%Paper: hep-th/9410213
%From: "David J. Fernandez C." <david@fis.cinvestav.mx>
%Date: Thu, 27 Oct 1994 14:48:56 -0700 (PDT)

\input harvmac.tex

\hfill CINVESTAV-FIS GFMR 11/94
\
\vskip1.5cm

\centerline{{\bf Geometric Phases and Mielnik's Evolution Loops}\footnote{$\sp
1$}{ Int. J. Theor. Phys. {\bf 33}, No. 10, 2037-2047 (1994)}}

\vskip1.5cm

\centerline{David J. Fern\'andez C.}

\centerline{\it Departamento de F\'{\i}sica, CINVESTAV-IPN}

\centerline{\it A.P. 14-740, 07000 M\'exico D.F., MEXICO}

\vskip1.5cm

\centerline{\bf Abstract.}

\bigskip\bigskip

The cyclic evolutions and associated geometric phases induced by
time-independent Hamiltonians are studied for the case when the evolution
operator becomes the identity (those processes are called {\it evolution
loops}). We make a detailed treatment of systems having equally-spaced energy
levels. Special emphasis is made on the potentials which have the same
spectrum as the harmonic oscillator potential (the generalized oscillator
potentials) and on their recently found coherent states.

\vfill

\noindent PACS: 03.65.Ca; 03.65.Sq

\eject

Since the appearance of Berry's work $\lbrack 1\rbrack$, much effort has been
spent in studying geometric aspects of nonrelativistic quantum mechanics
$\lbrack 2-8\rbrack$. In particular, to {\it any} cyclic evolution of the
vector state, $\vert \psi(\tau)\rangle = e\sp {i\phi}\vert \psi(0)\rangle$,
there has been associated a {\it geometric phase}
$$
\beta=\phi+\hbar\sp {-1} \int\sb 0\sp \tau \langle\psi(t)\vert  H(t) \vert
\psi(t)\rangle dt,  \eqno(1)
$$
where $\tau$ is the period of $\vert \psi(t)\rangle\in{\cal H}$,
$\langle\psi(t)\vert \psi(t)\rangle=1$, $\phi\in{\bf R}$, ${\cal H}$ is the
Hilbert space of vector states of the system, and $H(t)$ is the Hamiltonian
$\lbrack 3\rbrack$. $\beta$ describes global curvature effects arising on the
space of physical states, which is the projective space ${\cal P}$ formed by
the rays or the density operators $\vert \psi\rangle\langle\psi\vert $ instead
of ${\cal H}$. Due to this curvature, the horizontal lifting (parallel
transport) of the closed trajectory $\vert \psi(t)\rangle\langle\psi(t)\vert
\in{\cal P}$ leads to a trajectory $\vert \psi\sb H(t)\rangle$ which is, in
general, open on ${\cal H}$. The holonomy of this lifting is the Aharonov-
Anandan geometric phase factor $e\sp {i\beta}$.

The phase $\beta$, determined up to a multiple of $2\pi$, generalizes the
Berry phase, which originally was defined just for {\it adiabatic} cyclic
evolutions $\lbrack 1\rbrack$.  Further generalizations of $\beta$ have been
designed and can be found in the literature $\lbrack 3,5-6\rbrack$.  There is
also a lot of work dealing with the calculation of the geometric phases when
the Hamiltonian is time dependent (either explicitly or implicitly through
certain sets of time-dependent parameters $\lbrack 9-17\rbrack$). In this
paper, we will address the study of the geometric phases when the Hamiltonian
is time independent, i.e. $H(t)=H$. This choice is done because, it seems to
us, there is a widespread belief that the geometric phases appear only when
the Hamiltonian is time dependent, which is wrong. This is, perhaps, motivated
by the historical development of the subject and the following reasoning: the
eigenstates $\vert E\sb n\rangle$ of $H$ evolve according to $\vert E\sb
n(t)\rangle = e\sp {-iE\sb nt/\hbar}\vert E\sb n\rangle$, where the $E\sb n$
are the energy eigenvalues and $n$ denotes a set of discrete subscripts. These
evolutions are cyclic with period (arbitrary) $\tau$ and $\phi=-E\sb
n\tau/\hbar$. Therefore, from (1), $\beta=0$ for these states, and as usually
the only cyclic states at hand for these systems are the eigenstates of the
Hamiltonian, one is led to the wrong conclusion stated above. However,
Aharonov and Anandan found nonnull geometric phases for a spin $1/2$ in a
constant homogeneous magnetic field $\lbrack 3\rbrack$. The same will be true
for any other two-level system described by a time-independent Hamiltonian
$\lbrack 11,15\rbrack$. For nonspin systems it is possible to prove the
existence of nontrivial geometric phases for the harmonic oscillator $\lbrack
15,18\rbrack$ and some other physically interesting models (such as the
localized states of an electron on a crystal $\lbrack 15\rbrack$).

For independent reasons, in order to be used as the starting point for the
techniques of ``controling" and ``manipulating" the quantum systems, the {\it
evolution loops} (EL) were proposed (although without this name at that time)
in 1977 by Mielnik $\lbrack 19\rbrack$ and further developed by him in 1986
$\lbrack 20\rbrack$. Those loops are specific dynamical processes induced
either by time-dependent $\lbrack 19-21\rbrack$ or time-independent $\lbrack
22\rbrack$ Hamiltonians, for which the evolution operator $U(t)$ becomes the
identity ${\bf 1}$ (modulo a phase) for a certain time $\tau>0$, i.e.:
$$
U(\tau) = e\sp {i\phi}{\bf 1}, \eqno(2)
$$
where $U(0)={\bf 1}$ (see also $\lbrack 23\rbrack$). The EL are useful because
when perturbed by some additional external fields, the system can be driven to
attain any desired unitary operation on ${\cal H}$ due to the accumulation of
the small precessions of the distorted loop $\lbrack 19-21\rbrack$. In the
context of geometric phases a system performing an evolution loop is very
interesting because {\it any} state becomes cyclic at $t=\tau$:
$$
\vert \psi(\tau)\rangle = e\sp {i\phi}\vert \psi(0)\rangle. \eqno(3)
$$
Therefore, it could (we will show) have an associated nonnull geometric phase.
We will restrict ourselves in this paper to the evaluation of the geometric
phases associated to an evolution loop when the Hamiltonian is time
independent.

Suppose one has a system with a time-independent Hamiltonian $H$ whose
evolution operator performs an evolution loop. Hence, any vector state $\vert
\psi(t)\rangle$ comes back to itself at $t=\tau$ (see equations (2-3)), and
its geometric phase can be easily evaluated because the evolution operator
$U(t)=e\sp {-iHt/\hbar}$ commutes with $H$:
$$
\beta = \phi+\hbar\sp {-1} \int\sb 0\sp \tau \langle\psi(0)\vert U\sp
\dagger(t) H U(t)\vert \psi(0)\rangle dt= \phi + \hbar\sp {-1} \tau\langle H
\rangle , \eqno(4)
$$
where $\langle H \rangle=\langle\psi(0)\vert H\vert \psi(0)\rangle$.
Expressing $\vert \psi(0)\rangle$ in terms of the basis $\lbrace\vert E\sb
m\rangle\rbrace$, $\vert \psi(0)\rangle=\sum\sb mc\sb m\vert E\sb m\rangle$,
with $c\sb m=\langle E\sb m\vert \psi(0)\rangle$, we find that (4) becomes:
$$
\beta=\phi+\hbar\sp {-1}\tau\sum\sb m\vert c\sb m\vert \sp 2E\sb m.  \eqno(5)
$$

Note that formulas (4-5) are applicable to the cyclic evolution of a vector
state induced by any {\it time-independent} Hamiltonian regardless of whether
or not the system performs an evolution loop. However, if the system has an
evolution loop, then (4-5) will be valid for any initial condition. In
particular, for $\vert \psi(0)\rangle = \vert E\sb n\rangle$, i.e., $c\sb
m=\delta\sb {nm}$, it turns out that $\phi=-E\sb n\tau/\hbar$, and hence
$\beta=0$. If at least two $c\sb m$'s are distinct from zero, however, the
$\beta$ associated to the corresponding cyclic state will be, in general,
nontrivial (see $\lbrack 11\rbrack$, section 3.1).

There are in the literature some interesting systems whose time-indepen\-dent
Hamiltonian induces evolution loops $\lbrack 15,18,22-23\rbrack$. Here, we
will show the existence of an evolution loop for Hamiltonians whose spectrum
consists of equally spaced energy levels of the form:
$$
E\sb n=E\sb 0+n\Delta E, \eqno(6)
$$
where $\Delta E>0$ is the constant spacing between the levels and $E\sb 0$ is
the ground state energy. The subscript $n\in{\bf Z}\sp +$ takes values in
$\lbrack 0,N\rbrack $, where $N$ is finite if ${\cal H}$ is finite-dimensional
or infinite if ${\cal H}$ is infinite-dimensional. The evolution operator of
this system reads:
$$
U(t)=\sum\sb {n=0}\sp N e\sp {-iE\sb nt/\hbar}\vert E\sb n\rangle\langle E\sb
n\vert . \eqno(7)
$$
It is easy to see that the evolution loop is present at $\tau=2\pi\hbar/\Delta
E$:
$$
U(\tau)=\sum\sb {n=0}\sp N e\sp {-i2\pi(E\sb 0+n\Delta E)/ \Delta E}\vert E\sb
n\rangle\langle E\sb n\vert =e\sp {-i2\pi E\sb 0/\Delta E}{\bf 1}. \eqno(8)
$$
By comparing with (2), we obtain $\phi=-2\pi E\sb 0/\Delta E$. Moreover,
according to (4-5), the geometric phase for the cyclic state $\vert
\psi(t)\rangle$ is:
$$
\beta=2\pi{(\langle H\rangle -E\sb 0)\over\Delta E}=2\pi\sum\sb {n=1}\sp N
n\vert c\sb n\vert \sp 2\geq 0 .  \eqno(9)
$$
Notice that the component $c\sb 0$ of $\vert \psi(0)\rangle$ along the ground
state $\vert E\sb 0\rangle$ is not explicitly present in (9). If $\beta$ is
restricted (modulo $2\pi$) to the interval $\lbrack 0,2\pi)$, then equation (9)
admits the following interpretation: $\beta$ measures the ``energy excess" (in
dimensionless units) of $\langle H \rangle$ above the nearest lower energy
level $E\sb k$ (see Figure 1). If $E\sb k$ is given and $\vert \psi(0)\rangle$
is changed so that $E\sb k\leq\langle H \rangle <E\sb {k+1}$, then to the end
$\beta=0$ corresponds cyclic states with $\langle H \rangle=E\sb k$ (this
includes in particular $\vert \psi(0)\rangle = \vert E\sb k\rangle$). To any
other $\beta\in(0,2\pi)$ corresponds cyclic states with $\langle H \rangle
\neq E\sb k$ and vice versa (here necessarily $\vert \psi(0)\rangle\neq\vert
E\sb k\rangle$).

One of the interesting systems with equally spaced energy eigenvalues for
which our treatment can be applied is a spin $j$ interacting with a constant
homogeneous magnetic field ${\bf B}$, where $j>0$ can be either integer or
half-integer. Suppose, for simplicity, that the magnetic field points in the
$z$ direction, ${\bf B}=B{\bf k}$. The spin Hamiltonian can be expressed as
$H=-\mu{\bf J}\cdot{\bf B}=-\omega\sb cJ\sb 3$, where $\mu$ is the spin
magnetic moment, ${\bf J}$ is the spin operator whose components satisfy
$\lbrack J\sb k,J\sb l\rbrack = i \hbar\epsilon\sb {kln}J\sb n$, with
$k,l,n=1,2,3$, and $\mu B=\omega\sb c>0$ is the precession frequency of the
spin around ${\bf k}$. We work in the basis $\lbrace \vert j,m\rangle:J\sb
3\vert j,m\rangle=m\hbar\vert j,m\rangle, \ -j\leq m \leq j\rbrace$. Hence,
dim$({\cal H})=2j+1=N+1$. Due to the minus sign in the Hamiltonian, the
identifications $\Delta E=\hbar\omega\sb c$, $E\sb 0=-j\hbar\omega\sb c$, and
$\vert E\sb n\rangle=\vert j, j-n\rangle$ with $0\leq n\leq 2j=N$ are
consistent with equations (6-9). Therefore, the system performs an evolution
loop and so any spin state evolves in a cyclic way, with an associated
geometric phase given by equation (9). This is true, in particular, for the
spin $j=1/2$. In this case, it has become a convention to express the generic
initial state in terms of the spherical angles $\theta, \varphi$:
$$
\vert \psi(0)\rangle=e\sp {-i\varphi/2}\cos(\theta/2) \vert 1/2,1/2\rangle+
e\sp {i\varphi/2}\sin(\theta/2) \vert 1/2,-1/2\rangle.
$$
As the ground state in our notation is $\vert E\sb 0\rangle=\vert
1/2,1/2\rangle$, the only coefficient contributing to the geometric phase is
$c\sb 1=e\sp {i\varphi/2}\sin(\theta/2)$. Therefore, by applying (9), the
geometric phase becomes the usual one $\lbrack 3\rbrack$:
$$
\beta=2\pi\vert c\sb 1\vert \sp 2=2\pi\sin\sp 2(\theta/2)=\pi(1-\cos\theta).
$$
Moreover, as is well known, a problem involving just two energy levels can be
treated as a fictitious spin $1/2$ interacting with a homogeneous magnetic
field $\lbrack 11,15\rbrack$; hence, taking care in making a judicious
identification of the parameters, the same formulas to evaluate its geometric
phases may be applied.

At this point, it is worth discussing a geometric interpretation applicable to
systems with energy levels given by (6). It is easily understood for the
spin-$1/2$ system of the previous example, for which the space of physical
states (the projective space) coincides with the unit sphere $S\sp 2$ on ${\bf
R}\sp 3$. Any spin-$1/2$ state is precessing around the $z$ axis, performing
cyclic evolutions with a geometric phase which is, in general, distinct from
zero. There are two states, however, for which the evolution is trivial:
during the course of time they remain static at the north and south poles on
$S\sp 2$, and correspond to the spin aligned along and in the opposite
direction of the magnetic field. The geometric phase for both of them is zero.
For a system with $N+1$ equally-spaced energy levels, however, we have at hand
a more interesting (and more complicated) situation: now, instead of having
two static points on $S\sp 2$ there are $N+1$ points remaining static under
the evolution on ${\cal P}$ (those associated to the $\vert E\sb n\rangle$,
with $N$ either finite of infinite). For each one of them the geometric phase
is zero. Any other state will move across those points performing a more
complicated (cyclic) evolution on ${\cal P}$ with a nonnull geometric phase
(in general) which can be easily evaluated using equation (9).

We proceed now to the analysis of another system having equally-spaced energy
spectrum. It can be called the {\it generalized oscillator} (GO) because its
energy levels are exactly the same as the ones of the harmonic oscillator
Hamiltonian. The GO potentials were discovered by Abraham and Moses $\lbrack
24\rbrack$ using the Gelfand-Levitan formalism $\lbrack 25\rbrack$ and were
generated by Mielnik $\lbrack 26\rbrack$ using a generalization of the
well-known factorization method $\lbrack 27\rbrack$ (see also $\lbrack
28\rbrack$). Because of its didactic value, we will point out some steps used
in the generalized factorization to generate the GO potentials. We will work
from now on in the coordinate representation with dimensionless units
$\hbar=m=\omega=1$.

The {\it classical} factorization method applied to the oscillator consists in
expressing the Hamiltonian
$$
H={1\over 2}\left(-{d\sp 2\over dx\sp 2}+x\sp 2\right) \eqno(10)
$$
as the two products
$$
aa\sp \dagger=H+{1\over 2}, \ \ \ a\sp \dagger a=H-{1\over 2}, \eqno(11)
$$
where $a$ and $a\sp \dagger$ are the ordinary ladder operators $a =
(1/\sqrt{2})(d/dx+ x), \ \ a\sp \dagger=(1/\sqrt{2})(-d/dx+x)$ with $\lbrack
a, a\sp \dagger\rbrack ={\bf 1}$. The eigenfunctions and eigenvalues of the
harmonic oscillator can be constructed using the relations
$$
Ha\sp \dagger=a\sp \dagger(H+1), \ \ \  Ha=a(H-1). \eqno(12)
$$
There is a normalized ground state $\psi\sb 0(x)$ with eigenvalue $E\sb 0=1/2$
which satisfies $a\psi\sb 0(x)=0$ $\Rightarrow \ \psi\sb 0(x)\propto e\sp {-
x\sp 2/2}$, while the normalized eigenfunction $\psi\sb n(x)$  associated to
the eigenvalue $E\sb n=n+1/2$ is related to the ground state through:
$$
\psi\sb n(x)={(a\sp \dagger)\sp n\over \sqrt{n!}}\psi\sb 0(x).  \eqno(13)
$$

The {\it generalized} factorization method $\lbrack 26\rbrack$ consists in
looking for more general operators
$$
b={1\over\sqrt{2}}\left({d\over dx}+\beta(x)\right), \ \ \
b\sp \dagger={1\over\sqrt{2}}\left(-{d\over dx}+\beta(x)\right),  \eqno(14)
$$
satisfying just the first one of relations (11):
$$
bb\sp \dagger=H+{1\over 2}.  \eqno(15)
$$
Hence, the unknown function $\beta(x)$ obeys the Riccati equation
$$
\beta'+\beta\sp 2=1+x\sp 2,  \eqno(16)
$$
whose general solution is
$$
\beta(x)=x+{e\sp {-x\sp 2}\over \lambda+\int\sb 0\sp x e\sp {-y\sp 2} dy}, \ \
\lambda\in{\bf R}. \eqno(17)
$$
Now, the point is that the product $b\sp \dagger b$ is no longer related to
the harmonic oscillator Hamiltonian, but it leads to a new operator $H\sb
\lambda$:
$$
b\sp \dagger b=H\sb \lambda-{1\over 2},  \eqno(18)
$$
where
$$
H\sb \lambda =-{1\over 2}{d\sp 2\over dx\sp 2}+V\sb \lambda(x), \eqno(19)
$$
with
$$
V\sb \lambda(x)={x\sp 2\over 2}-{d\over dx}\left({e\sp {-x\sp 2} \over
\lambda+\int\sb 0\sp x e\sp {-y\sp 2}dy}\right)=\left(x+{e\sp {-x\sp
2}\over\lambda+\int\sb 0\sp x e\sp {-y\sp 2}dy}\right)\sp 2 -{x\sp 2\over 2}.
\eqno(20)
$$
The requirement $\vert \lambda\vert >\sqrt{\pi}/2$ assures that $V\sb
\lambda(x)$ has no singularities. The relationships analogous to (12) provide
the way to obtain the eigenfunctions and eigenvalues of $H\sb \lambda$:
$$
H\sb \lambda b\sp \dagger=b\sp \dagger(H+1), \ \ \ Hb=b(H\sb \lambda -1).
\eqno(21)
$$
Hence, the states $\theta\sb n(x)=b\sp \dagger \psi\sb {n-1}(x)/\sqrt{n}, \
n=1,2,\cdots$, are orthonormalized eigenfunctions of $H\sb \lambda $ with
eigenvalues $E\sb n=n+1/2$. However, the set $\lbrace\theta\sb n(x), \ n= 1,2,
\cdots\rbrace$ is not yet a basis of $L\sp 2({\bf R})$. There is a missing
unit vector $\theta\sb 0(x)$ which is orthogonal to all the vectors $\theta\sb
n(x), n=1,2,\cdots$. It turns out to be an eigenfunction of $H\sb \lambda $
with eigenvalue $E\sb 0=1/2$ satisfying $b\theta\sb 0(x)=0$, and taking the
form:
$$
\theta\sb 0(x)\propto\exp\left(-\int\sb 0\sp x \beta(y)dy\right).  \eqno(22)
$$
As the set $\lbrace\theta\sb n(x), \ n= 0,1,2, \cdots\rbrace$ forms a basis in
$L\sp 2({\bf R})$, then $\lbrace H\sb \lambda : \vert \lambda\vert
>\sqrt{\pi}/2\rbrace$ is a family of Hamiltonians distinct from the harmonic
oscillator one but which has exactly the same spectrum as the oscillator has.
In the limit $\vert \lambda\vert \rightarrow \infty$, the harmonic oscillator
potential is recovered, $V\sb \lambda(x)\rightarrow x\sp 2/2$ when $\vert
\lambda\vert \rightarrow\infty$.

All the relationships involving the evolution loops and the geometric phase
(equations (6-9)) can be applied to the GO Hamiltonian (19-20) with $E\sb
0=1/2, \ \Delta E=1, \ \tau=2\pi, \ \phi=-\pi$, and $N=\infty$. In particular,
the geometric phase is $\beta=2\pi(\langle H\sb \lambda  \rangle -1/2)$, and
when applied to the wavefunctions of the basis $\lbrace\theta\sb n(x),
n=0,1,2\cdots\rbrace$ we recover again $\beta=2n\pi$. Is there any other set
of generic states of the GO potential for which we can evaluate explicitly the
geometric phase? The answer turns out to be positive when considering the
family of recently found coherent states for the GO Hamiltonian $\lbrack
29\rbrack$. Here, we will present some details of its derivation (for work
involving coherent states and their geometric phases see $\lbrack
9,11,15,18,30-32\rbrack$).

In the construction of the coherent states of $H\sb \lambda$, denoted as
$\vert z\rangle$ with $z\in{\bf C}$,  we need to identify the ``annihilation"
and ``creation" operators of the system. Because $b\theta\sb
n(x)\propto\psi\sb {n-1}(x) \ \Rightarrow \ ab\theta\sb n(x)\propto\psi\sb {n-
2}(x) \ \Rightarrow \ b\sp \dagger ab\theta\sb n(x)\propto \theta\sb {n-
1}(x)$, and an obvious choice is:
$$
A=b\sp \dagger ab, \ \ \ A\sp \dagger=b\sp \dagger a\sp \dagger b. \eqno(23)
$$
The coherent states can be defined now as the eigenstates of the annihilation
operator $A$ with eigenvalues $z$, i.e. $A \vert z\rangle=z\vert z\rangle$.
Expressing $\vert z\rangle$ in terms of the basis $\lbrace\vert \theta\sb
n\rangle, \ n=0,1,2\cdots\rbrace$, and substituting explicitly that
expression in the previous one, we find the following family of coherent
states (after normalization):
$$
\vert z\rangle={1\over\sqrt{{}\sb 0F\sb 2(1,2;\vert z\vert \sp 2)}}\sum\sb
{n=0}\sp \infty {z\sp n\over n!\sqrt{(n+1)!}}\vert \theta\sb {n+1}\rangle ,
\eqno(24)
$$
where $\vert \theta\sb n\rangle$ is the ket representing the eigenfunction
$\theta\sb n(x)$ and ${}\sb 0F\sb 2(1,2;y)$ is a generalized hypergeometric
function defined by:
$$
{}\sb 0F\sb 2(\alpha,\beta;y)=\sum\sb {n=0}\sp
\infty{\Gamma(\alpha)\Gamma(\beta)\over \Gamma(\alpha+n)\Gamma(\beta+n)} {y\sp
n\over n!}, \eqno(25)
$$
with $\Gamma(\cdot)$ the Gamma function. To each value $z\neq 0$ corresponds
one and only one coherent state. However, $z=0$ is a doubly degenerate
eigenvalue of $A$ with two orthogonal eigenvectors which will be denoted
$\vert \theta\sb 0\rangle$ and $\vert z=0\rangle=\vert \theta\sb 1\rangle$. By
choosing an appropiate measure in the complex plane, it can be shown that the
set $\lbrace\vert \theta\sb 0\rangle, \vert z\rangle\rbrace$ is complete in
${\cal H}$.

The relationships presented so far are sufficient for our purpose of
evaluating the geometric phase. To this end, we need to find the expected
value of $H\sb \lambda$ in the state $\vert z\rangle$. A direct calculation
leads to:
$$
\langle H\sb \lambda \rangle=\langle z\vert H\sb \lambda \vert
z\rangle=1/2+{{}\sb 0F\sb 2(1,1;\vert z\vert \sp 2)\over {}\sb 0F\sb
2(1,2;\vert z\vert \sp 2)}.  \eqno(26)
$$
Finally, substituting (26) in the equation for $\beta$, we obtain the
following expression for the geometric phase $\beta\sb {GCS}$ of the
generalized coherent state:
$$
\beta\sb {GCS}=2\pi{{}\sb 0F\sb 2(1,1;\vert z\vert \sp 2)\over
{}\sb 0F\sb 2(1,2;\vert z\vert \sp 2)}.  \eqno(27)
$$
To have an idea of the behaviour of $\beta\sb {GCS}$, we plot it versus ${\rm
Re}(z)\times{\rm Im}(z)$ in Figure 2. As we can see, the geometric phase is
independent of $\lambda$ and depends on $z$ in a quite different way compared
with that of a standard coherent state (SCS) of the harmonic oscillator, for
which $\beta\sb {SCS}=2\pi\vert z\vert \sp 2$ $\lbrack 15,18\rbrack$ (see also
Figure 2). This occurs because the generalized coherent states discussed in
$\lbrack 29\rbrack$ do not tend to the standard ones when
$\lambda\rightarrow\infty$ even though the generalized potential tends to the
harmonics oscillator potential in this limit. A deeper analysis shows that the
difference rests on the fact that in this limit the annihilation operator
$A\sb \infty\equiv\lim\sb {\lambda\rightarrow\infty} A=a\sp \dagger a\sp 2$ is
distinct from the standard one $a$. The generalized coherent states, however,
could be useful in future applications because the product of the uncertainty
of the $\hat X$ and $\hat P$ operators for these states is almost minimum in
this limit $1/2\leq\lim\sb {\vert \lambda\vert \rightarrow\infty}\Delta {\hat
X}\Delta{\hat P}\leq 3/2$. The question of whether or not there is a family of
coherent states of $H\sb \lambda$ tending to the standard ones when
$\lambda\rightarrow\infty$, the geometric phases included, is open.

\bigskip

The author acknowledges CONACYT (M\'exico) for financial support.

\vfill\eject

\centerline{\bf References.}

\bigskip

\item{1.} M.V. Berry, Proc. R. Soc. Lond. A {\bf 392}, 45 (1984).

\item{2.} B. Simon, Phys. Rev. Lett. {\bf 51}, 2167 (1983).

\item{3.} Y. Aharonov and J. Anandan, Phys. Rev. Lett. {\bf 58}, 1593 (1987).

\item{4.} J. Anandan, Ann. Inst. Henri Poincar\'e {\bf 49}, 271 (1988).

\item{5.} J. Samuel and R. Bhandari, Phys. Rev. Lett. {\bf 60}, 2339 (1988).

\item{6.} J. Anandan and Y. Aharonov, Phys. Rev. Lett. {\bf 65}, 1697 (1990).

\item{7.} A. Bohm, L.J. Boya and B. Kendrick, Phys. Rev. A {\bf 43}, 1206
(1991).

\item{8.} L.J. Boya, J.F. Cari\~nena and J.M. Gracia-Bond\'{\i}a, Phys. Lett.
A {\bf 161}, 30 (1991).

\item{9.} E. Layton, Y. Huang and S.I. Chu, Phys. Rev. A {\bf 41}, 42 (1990).

\item{10.} D.J. Moore and G.E. Stedman, J. Phys. A {\bf 23}, 2049 (1990).

\item{11.} D.J. Moore, Phys. Rep. {\bf 210}, 1 (1991).

\item{12.} D.J. Fern\'andez C., L.M. Nieto, M.A. del Olmo and M. Santander, J.
Phys. A {\bf 25}, 5151 (1992).

\item{13.} D.J. Fern\'andez C., M.A. del Olmo and M. Santander, J. Phys. A
{\bf 25}, 6409 (1992).

\item{14.} D.J. Fern\'andez C. and N. Bret\'on, Europhys. Lett. {\bf 21}, 147
(1993).

\item{15.} A.N. Seleznyova, J. Phys. A {\bf 26}, 981 (1993).

\item{16.} J.C. Solem and L.C. Biedenharn, Found. Phys. {\bf 23}, 185 (1993).

\item{17.} P. Campos, J.L. Lucio M. and M. Vargas, Phys. Lett. A {\bf 182},
217 (1993).

\item{18.} M.G. Benedict and W. Schleich, Found. Phys. {\bf 23}, 389 (1993).

\item{19.} B. Mielnik, Rep. Math. Phys. {\bf 12}, 331 (1977).

\item{20.} B. Mielnik, J. Math. Phys. {\bf 27}, 2290 (1986).

\item{21.} D.J. Fern\'andez C. and B. Mielnik, J. Math. Phys. {\bf 35},
2083 (1994).

\item{22.} D.J. Fern\'andez C., Il Nuovo Cimento {\bf 107 B}, 885 (1992).

\item{23.} M.M. Nieto and V.P. Gutschick, Phys. Rev. D {\bf 23}, 922 (1981).

\item{24.} P.B. Abraham and H.E. Moses, Phys. Rev. A {\bf 22}, 1333 (1980).

\item{25.} I.M. Gelfand and B.M. Levitan, Am. Math. Soc. Transl. {\bf 1}, 253
(1951).

\item{26.} B. Mielnik, J. Math. Phys. {\bf 25}, 3387 (1984).

\item{27.} L. Infeld and T.E. Hull, Rev. Mod. Phys. {\bf 23}, 21 (1951).

\item{28.} D.J. Fern\'andez C., Lett. Math. Phys. {\bf 8}, 337 (1984).

\item{29.} D.J. Fern\'andez C., V. Hussin and L.M. Nieto, J. Phys. A {\bf 27},
3547 (1994).

\item{30.} G. Giavarini and E. Onofri, J. Math. Phys. {\bf 30}, 659 (1989).

\item{31.} Y. Brihaye, S. Giler, P. Kosi\'nski and P. Ma\'slanka, J. Phys. A
{\bf 23}, 1985 (1990).

\item{32.} M. Maamache, J.P. Provost and G. Vall\'ee, J. Phys. A {\bf 23},
5765 (1990).

\vfill\eject

\noindent{\bf Fig.1} Schematic representation of the $N+1$ energy levels for a
system with equally-spaced spectrum. If the geometric phase $\beta$ is
restricted to the interval $\lbrack 0,2\pi)$, then it can be interpreted as an
energy excess of the system with respect to its nearest energy level $E\sb k$
(the nearest below) in dimensionless units.

\bigskip

\noindent{\bf Fig.2} The geometric phases associated to the standard coherent
states of the harmonic oscillator ($\beta\sb {SCS}$) and the coherent states
of the generalized oscillator ($\beta\sb {GCS}$) as functions of the complex
variable $z$. The minimum values of $\beta\sb {SCS}$ and $\beta\sb {GCS}$ are
$0$ and $2\pi$ respectively, both at $z=0$. The missing sections in both
surfaces were removed to show the behaviour close to the minimum.

\end